\documentclass[10pt,journal,compsoc]{IEEEtran}

\usepackage{amsmath,amssymb,amsfonts}
\usepackage[linesnumbered, ruled, vlined]{algorithm2e}
\usepackage{graphicx}
\usepackage{xcolor}
\usepackage{multirow}

\usepackage{textcomp} 
\newcommand{\qdist}[1]{\ifmmode\langle#1\rangle\else\textlangle#1\textrangle\fi}

\ifCLASSOPTIONcompsoc
  \usepackage[nocompress]{cite}
\else
  \usepackage{cite}
\fi

\ifCLASSINFOpdf
\else
\fi

\begin{document}

\title{On Designing GPU Algorithms with Applications to Mesh Refinement}

\author{Zhenghai Chen,
        Tiow-Seng Tan,
        and Hong-Yang Ong 
\IEEEcompsocitemizethanks{
\IEEEcompsocthanksitem Authors are with the School of Computing, National University of Singapore. 
\IEEEcompsocthanksitem Emails: \{dcschai $|$ dcstants $|$ dcsohy\}@nus.edu.sg.
}
}


\IEEEtitleabstractindextext{%
\begin{abstract}
We present a set of rules to guide the design of GPU algorithms. These rules are grounded on the principle of reducing waste in GPU utility to achieve good speed up. 
In accordance to these rules, we propose GPU algorithms for 2D constrained, 3D constrained and 3D Restricted Delaunay refinement problems respectively. Our algorithms take a 2D planar straight line graph (PSLG) or 3D piecewise linear complex (PLC) $\mathcal{G}$ as input, and generate quality meshes conforming or approximating to $\mathcal{G}$.
The implementation of our algorithms shows that they are the first to run an order of magnitude faster than  current state-of-the-art counterparts in sequential and parallel manners while using similar numbers of Steiner points to produce triangulations of comparable qualities. It thus reduces the computing time of mesh refinement from possibly hours to a few seconds or minutes for possible use in interactive graphics applications.
\end{abstract}

}

\maketitle

\IEEEdisplaynontitleabstractindextext

\IEEEpeerreviewmaketitle

\begin{figure*}[t]
 \centering
 \includegraphics[scale = 0.2]{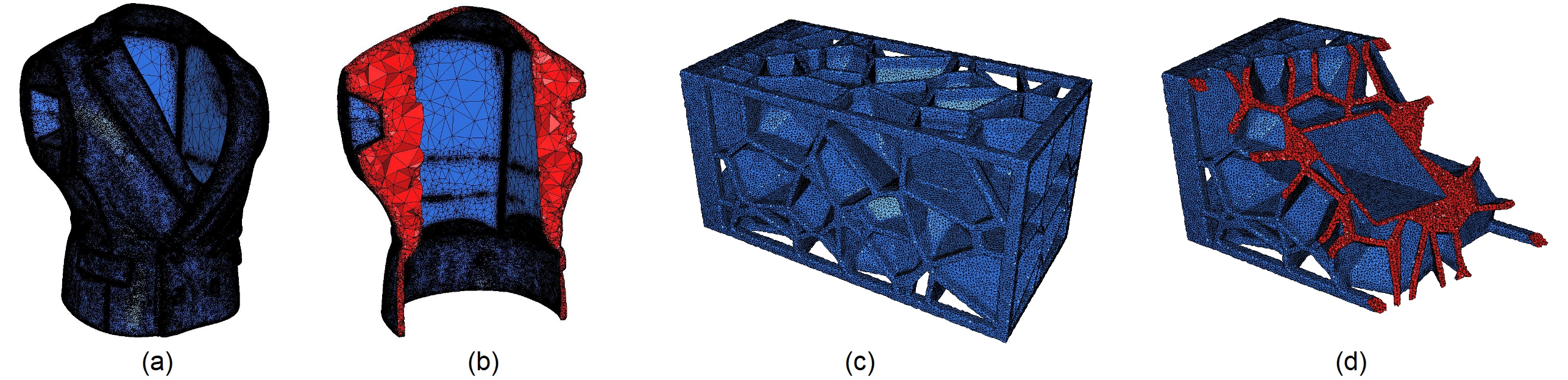}
 \vspace{-0.1in}
 \caption{Meshes generated by our GPU algorithms. 
 (a) and (b) 3D quality CDT and its cut-off view generated by {\tt gQM3d} on the Model 763718, Jacket.
 (c) and (d) 3D quality RDT and its cut-off view generated by {\tt gDP3d} on the Model 940414, Voronoi Lamp.
 }
 \label{fig:teaser}
\vspace{-0.1in}
\end{figure*}

\IEEEraisesectionheading{\section{Introduction}\label{sec:introduction}}

\IEEEPARstart{F}{or} more than a decade, the graphics processing unit (GPU) has been used in general-purpose computing. 
Many models and methodologies in general parallel computing have been adapted into the studies of  GPU algorithms.
These include PRAM models (see \cite{JaJa92}), CTA model (see \cite{LinSnyder2008}), the PCAM design process \cite{PCAM95}, methodologies for designing parallel and multi-threading applications \cite{Quinn86, Breshears09} and so on.    

One area, however, has not been well addressed, to the best of our knowledge.  A good GPU program running on a GPU with many cores (thousands as of now, potentially scaling up greatly in the future), may not necessarily do only \textit{useful} computations, that is, computations which directly contribute to, or are necessary to achieve, the final solution. This is because the efforts to identify these work may require additional work, and/or cause under-utilization of the hardware resources, potentially resulting in lowered overall performance.  In some cases, doing \textit{speculative} computations, that is, computations which may not contribute to, or be necessary to achieve, the final solution, can achieve better overall performance, due to better utilization of the hardware.  As such, this \textit{waste}, i.e., computations that are not {useful}, can be seen as a necessary evil in many contexts, and is orthogonal to concepts such as work regularization and data localization, among others commonly considered as good practices in GPU computing.

For example, in the mesh refinement \cite{ChengBook2002Meshing} which requires the addition of Steiner vertices, the amount of necessary computation required cannot be known beforehand.  Depending on the specific problem set, the amount of computations can range from very high (when there are many available Steiner points to be added at that instance) to very low (when there are few Steiner points to be added). In addition, waste can result from abandoned results due to dependency issues or under-utilization of computing resources.
This contrasts greatly with problems such as matrix operations, where the amounts of computation or flow of control are relatively static or highly predictable, and any waste directly implies a penalty in overall performance.

There are many other interesting computational problems that have this characteristic similar to mesh refinement. 
The algorithms for solving this type of problems can often be expressed at high-level as \textit{iterations} of computation until a \textit{terminating condition} is reached.  This is the main kind of problems and algorithms that this paper is targeted at, when applied to GPU computing.

Though there are already many references on good practices and optimizations \cite{CUDA19}, as well as  understanding and prediction of the performance 
(see 
\cite{Volkov:EECS-2016-143} 
and references therein) 
for the GPU, all these are targeted at GPU programs at a low level. 
To this end, the paper makes the following contributions: 

\begin{enumerate}

\item 
A set of simple rules (or strategies) to manage and optimize waste against overall performance based on prediction of workload and computational resource utilization to achieve better speed up.

\item 
A case study using GPU algorithm for 2D constrained Delaunay mesh refinement ({\tt gDP2d}) showing an order of magnitude speed up compared to the CPU algorithm in {\tt Triangle} \cite{Shewchuk96}.

\item
A case study using GPU algorithm for 3D constrained Delaunay mesh refinement ({\tt gQM3d}) showing an order of magnitude speed up compared to the CPU algorithm in {\tt TetGen} \cite{Si15}; see Figure~\ref{fig:teaser}(a) and (b).

\item 
A case study using GPU algorithm for 3D restricted Delaunay mesh refinement ({\tt gDP3d}) showing an order of magnitude speed up compared to the multi-threading CPU algorithm in {\tt CGAL} \cite{cgal3d}; see Figure~\ref{fig:teaser}(c) and (d).

\end{enumerate}

Section~\ref{sec:designRules} below describes the set of simple rules. Section~\ref{sec:meshPreliminary} provides background on the mesh refinement problems. Section~\ref{sec:2D}, Section~\ref{sec:3D-CDT}, and Section~\ref{sec:3D-RDT} 
are the case studies for the 2D constrained, 3D constrained and 3D restricted Delaunay mesh refinement problems respectively,
with Section~\ref{sec:experiment} showing comparisons in the resultant performances, leading to concluding remarks in Section~\ref{sec:conclusion}.

\section{Design Rules on GPU Computation}
\label{sec:designRules}

We start by assuming that the computational problem is already partitioned into parallel {tasks} via, for example, the PCAM process \cite{PCAM95}. At this point, it is assumed that it is already clear how the data structures and the kernel programs, hereafter simplified as {\em kernels}, are developed into the basic GPU algorithm to solve one iteration of the computational problem.  The computation required for checking the terminating condition is assumed to remain unchanged, and will be ignored in this context.
This GPU algorithm will then be augmented with additional modifications, guided by our design rules, to try to optimize the performance, mainly based on the workload situation.
When running a kernel over one iteration, the GPU is said to be under \textit{low workload} when the occupancy is low, or \textit{high workload} when the occupancy is high.

A kernel can launch many threads at once, where each thread can be assigned to perform some computation with respect to some unit of work, or simply \textit{work}, in parallel with other threads.  Naively, the more threads a kernel can launch at once, the more work can be computed per launch, thus resulting in faster completion.  However, the maximum number of threads is restricted by the actual GPU hardware, primarily the number of processing cores and the total amount of memory available compared to the amount of local memory required by the work in the threads. For simplicity's sake, we assume all work assigned to threads in a launch must be completed before another kernel can launch new threads.  
The duration from the time when the threads are launched to the time the slowest thread finished its work is the \textit{latency} for the launch.  And the total number of useful work done for the launch is the \textit{concurrency}.  By Little's Law \cite{Little61}, the \textit{throughput} for the launch is the ratio of concurrency to the latency.  This throughput is used as a  momentary performance indicator for the GPU algorithm where high throughput can be achieved by either increasing the work done or decreasing the latency.

\subsection*{\underline{Design Rules to Reduce Waste}}

Our main design principle lies in managing waste reduction efforts.
Waste can occur due to under-utilization of the GPU, i.e., not using up to the maximum available number of threads for the current work.  It can also happen when completed or in-progress work is invalidated due to some reasons, such as detecting a dependency conflict either during, or after the thread has completed its work. In general, waste contributes negatively to overall performance.
On the other hand, we can perform pre-computations or incorporate additional steps within the GPU algorithm to attempt to reduce the waste mentioned above. But these usually come at a cost of their own and there may be cases where it is not worth the effort to attempt waste reduction when the cost is high compared to the potential savings from the 
reduced resultant waste.
To address these concerns, we propose the following high-level guidelines, applicable to a general GPU algorithm, to potentially achieve better overall performance.

\begin{description}

    \item [{\bf Rule 1:}] {When workload is low, forgo filtering out threads that do no useful work.}
   
    \item [{\bf Rule 2:}] {When workload is high, attempt to shorten latency and/or avoid scheduling work that are likely to result in dependency conflict.}
  
    \item [{\bf Rule 3:}] {When workload shifts from high to low in a kernel, consider terminating threads still running when most other threads have completed.}
    
    \item [{\bf Rule 4:}] {To prevent interspersed iterations with low workload from affecting overall performance negatively, consider merging planned work from the next iteration (as a form of speculative computation) to potentially achieve better computational resource utilization.}
    
    \item [{\bf Rule 5:}] {To prevent a long run of iterations with very low workload and long latency, consider breaking up lengthy work to be completed over multiple iterations to improve the throughput.}
    
\end{description}

{\bf Rule~1}, {\bf Rule~2} and {\bf Rule~3} apply to workload scenarios within one kernel's launching of threads to do work. 
{\bf Rule~4} and {\bf Rule~5} are applied to a serial of iterations within the algorithm at a higher level. 
For our purposes, we do not need to define precisely what constitute an iteration, but use it loosely to mean executing a whole computational loop to, for example, insert Steiner points in a mesh refinement problem, or executing a bundle of kernels which is only a part of the computational loop. 
We next look at each rule in detail.

\subsection*{\underline{Rule 1. No waste when in low workload}}

In GPU programming, compaction (e.g., using parallel prefix) is an efficient technique commonly used to filter out works which are known to be useless so that they will not be assigned to threads, thus potentially achieving greater concurrency. On the other hand, this incurs additional computational costs, which can be represented by time duration $\ell$ added to the original latency ${L}$.  

However, when workload is low, it can be assumed that there are no pending work at that moment which has not already been assigned to available threads, i.e., there is no work to be assigned even if more threads are made available.  Thus, there is no change to the concurrency $C$. This results, by Little's Law, in a final throughput $\frac{C}{{L} + \ell}$ lower than $\frac{ C}{{L}}$ when no compaction is done.  Thus, in this scenario, it is better not to attempt any compaction at all.

\subsection*{\underline{Rule 2. Reducing waste when high workload}}

Given some work greater than or equal to the number of available threads within one iteration of a GPU program, the ideal case would be for all the work to be of short latency and independent from one another and all the threads to be assigned useful work in each launch (except maybe the last), thus achieving optimal performance.
In practice, however, waste can happen due to dependency in the work, which can cause completed work to be invalidated, rolled back and re-attempted later, upon detection of conflicts.


If the waste is significant, which is usually the case with high workload, it may be worthwhile to perform additional computation to filter the work to reduce potential conflicts.
Let the ideal concurrency of the GPU be ${C}$ and the effective concurrency due to waste be $\alpha {C}$ where $0 < \alpha < 1$. Then the effective throughput is ${T} = \frac{\alpha {C}}{L}$ by Little's Law where ${L}$ is the latency. 
Suppose that we perform filtering before assigning the work. We improve the effective concurrency to $\beta {C}$, where $\alpha < \beta \leq 1$, but increased latency from $L$ to ${L} + \ell$ where $\ell$ is the additional computational time due to the filtering. For the effective throughput to be larger than $T$ while employing filtering, we require $\frac{{ L}+\ell}{L} < \frac{\beta}{\alpha}$. That is, the proportion of time needed to do the filtering must be lower than the proportion of increase in the effective concurrency. This can be used as an indicator for testing the performance of specific filtering method in empirical studies.

\subsection*{\underline{Rule 3. Reducing waste from lengthy work}}

Threads within a launch may take different amounts of time to complete their work.  When this happens, threads which have already completed their work are forced to wait for threads which are still processing, resulting in waste.  This can be due to inappropriate high-level algorithm and data structure design or simply intrinsic to the specific problem.  When most threads have completed their work, we want to consider terminating the remaining threads early so as to reduce the effective latency (reducing waste due to 
threads with no more work), 
at the cost of additional waste from the abandoned work.

Let $L$ be latency in the original case where all completed threads wait for the slowest thread to complete, where $C$ is the effective concurrency, giving the throughput $T = \frac{C}{L}$ by Little's Law.
Suppose we terminate the remaining threads after a fraction $(1-\gamma)$ of the threads have completed their work at $({L} - \ell)$ since launch.  Assuming that the terminated threads are evenly distributed with respect to effective concurrency, we can achieve a mean throughput of $\frac{(1 - \gamma)  C}{{L} - \ell}$. For there to be improvement in performance, $\frac{(1 - \gamma) C}{{L} - \ell} > \frac{C}{L}$ , we need $\ell > \gamma L$, where $\ell$ is the reduction in latency. That is, the saving in latency should be larger than the loss in concurrency.

This, however, is insufficient in the determination of a good value for $\gamma$ since $L$ is unknown.  In practice, a simple implementation is to simply assign $\gamma$ to be 20\%, from the Pareto Principle a.k.a. 80:20 rule. A more advanced method is to monitor the rate of thread completion and choose to terminate remaining threads when the rate drops significantly after a certain point where most threads have completed their work. The work in terminated threads can be re-attempted in subsequent iterations.

\subsection*{\underline{Rule 4. Reducing waste in fluctuating workload}}

The high-level algorithm often expresses required computations in terms of groups of data and operations over them. Naively, one group of data and its operations translate to one iteration in the GPU algorithm, which will then perform the operations on elements in the group concurrently in threads.
In many algorithms, 
it is observed that there can be great variance in the group sizes, often pattern-like, that flips between large and small groups.  This causes the running GPU program to have iterations that fluctuate between high and low workload, negatively affecting performance.  One way to improve this is to see whether we can merge the work from a small group with other groups by unifying their operations.  This may give better utilization at the cost of increased latency due to the more complex kernel.

Let ${T}_i$, ${L}_i$ and ${C}_i$ be the throughput, latency and concurrency for work in group $i$, 
and ${T}_j$, ${L}_j$ and ${C}_j$ be those respectively for group $j$.
The throughputs are ${T}_i = \frac{{ C}_i}{{L}_i}$ and ${T}_j = \frac{{ C}_j}{{L}_j}$ respectively by Little's Law. 
Assume we merge the work in the two groups so that they are processed concurrently, with resultant concurrency ${C}_m$ and latency ${L}_m$.
It is expected that ${C}_m \ge \max({C}_i, {C}_j)$ and ${L}_m \ge \max({L}_i, {L}_j)$ or ${L}_m = \max({L}_i, {L}_j) + \ell$, where $\ell \ge 0$.
Compare the expected throughput where we process the two groups concurrently ${T}_m = \frac{{C}_m}{{L}_m}$, against the throughput if we process the groups in series ${T}_s = \frac{{C}_i + {C}_j}{{L}_i+{L}_j}$.
In the optimal case where ${C}_m \approx {C}_i + {C}_j$, we can achieve better throughput, i.e., ${T}_m > {T}_s$, if $\ell < \min({L}_i, {L}_j)$. In the less than optimal case, however, $\ell$ may have to be smaller to achieve improvement, or there may even be cases where improvements are not possible due to low ${C}_m$ or high $\ell$. The trade-off can be determined empirically.

It is noted that the state-of-the-art GPU is capable of utilizing spare capacity left over from a kernel to launch other kernels to do computations under some constraints. 
We disregard this for our discussion here. 

\subsection*{\underline{Rule 5. Reducing waste in serial of low workload}}

Long runs of iterations with very low workload cause lots of waste, due to under-utilization of the GPU over long duration. When this cannot be addressed well by {\bf Rule 4}, 
we can always consider falling back to using the CPU to complete the work
instead, 
which may give better performance.
This is particularly true when this scenario happens during the starting or the ending of the algorithm. On the other hand, when this occurs while the algorithm is in the middle of its computation, falling back to the CPU may not be feasible due to the relatively high cost of data transfer between GPU and CPU.    

In the latter case, one can consider breaking up work to be done over multiple iterations to shorten latency for any one iteration. This then involves carrying forward useful data or information from an iteration to be processed in the next iteration and so on. 
This is at the expense of lower concurrency for an iteration (because fewer work are completed), but not so for a serial of iterations since useful data are kept from iteration to iteration to still complete all the work in subsequent iterations. 
Thus, the additional cost here is the computation to record useful data for subsequent reuses.
Similar to those arguments in the above rules using Little's Law, 
one can obtain a better throughput when the saving in computational time from shortening latency in the serial of low workloads is more than enough to justify the increase in computational cost.

\section{Preliminary on Mesh Refinement}
\label{sec:meshPreliminary}

Delaunay mesh, a mesh satisfying Delaunay property \cite{ChengBook2002Meshing}, is widely used in various computer graphics, computer aided, engineering and scientific applications. Such applications usually require meshes, besides being Delaunay, to meet certain quality criteria such as bounds on the triangle area, angle, edge length, tetrahedron volume, aspect ratio, etc. 
The quality of a Delaunay mesh is improved by inserting additional points, \emph{Steiner points}, into the mesh, through a process known as \emph{Delaunay mesh refinement}. Substantial efforts have been invested in sequential algorithms to generate and refine Delaunay meshes with constraints on segments and triangular faces \cite{Shewchuk96, cgal2d, cgal3d}. However, an increase in the number of input constraints or a stricter criterion can incur a very huge cost in required time for sequential algorithms. There is thus a desire to explore parallel computing with respect to the Delaunay mesh refinement problem.

With the GPU, an 
approach is to simply mimic the sequential approach but allow Steiner points to be inserted concurrently in each iteration and then terminate once the resultant mesh satisfies all the quality conditions. 
However, this cannot guarantee improvement to the performance, due to unbalanced workloads and dependency conflicts.  In addition, it introduces a new problem where superfluous points added may seriously affect the termination of the algorithm,  producing an output mesh of an unnecessary large size.   The situation is further complicated by the additional constraints on edges, triangles or tetrahedra, where specialized computations have to be done in a certain order.


In accordance to the design rules, 
the following three sections present our mesh refinement algorithms in GPU: {\tt gDP2d} (Section~\ref{sec:2D}), {\tt gQM3d} (Section~\ref{sec:3D-CDT}), and {\tt gDP3d} (Section~\ref{sec:3D-RDT}). Interested reader may refer to 
\cite{Chen17, Chen19, Chen:Thesis} and other references provided therein 
for background details.
%
%
%
%

\section{2D Constrained Delaunay} 
\label{sec:2D}

A \emph{planar straight line graph} (PSLG) $\mathcal{G}$ in 2D contains a set of vertices $P$ and a set of non-crossing edges $E$ where each is formed with endpoints in $P$. 
A {\em triangulation} $\mathcal{T}$ of $\mathcal{G}$ is a decomposition of the convex hull of $P$ into triangles with all vertices and edges of $\mathcal{G}$ appearing as vertices and union of edges 
in $\mathcal{T}$. In subsequent discussion, we call an edge in $E$ a {\em segment} while an edge in $\mathcal{T}$ which is also a part or whole of some segment a {\em subsegment}. Then,   
a constrained Delaunay triangulation (CDT) for $\mathcal{G}$ is a triangulation $\mathcal{T}$ such that the circumcircle of each of its triangle $t$ does not contain any vertices except for possibly those not visible (as blocked by edges in $E$) to all three vertices of $t$.
If the mentioned exception is dropped, the CDT is also a so-called {\em conforming Delaunay triangulation}. To simplify our discussion, we use just CDT which is possibly a conforming Delaunay triangulation too.

\subsection{2D CDT Problem}

Given an input PSLG $\mathcal{G}$, angle $\theta$ and edge length $l$, the goal is to output a CDT $\mathcal{T}$ of $\mathcal{G}$ that contains few, if not zero, \emph{bad} triangles. A triangle is said to be \emph{bad} if it has an angle smaller than $\theta$ or an edge longer than $l$. If there are some input angles, formed by segments in $\mathcal{G}$, smaller than $60^\circ$, the algorithm will end up with some bad triangles near the small input angles \cite{Shewchuk00}. 
Such bad triangles are unavoidable and are accepted to be in the output \cite{Ruppert95, Shewchuk97}. 
The open source sequential program {\tt Triangle} by Shewchuk \cite{Shewchuk96} and {\tt CGAL} 2D mesh generator \cite{cgal2d,Boissonnat02} are widely used for the purpose.
The former implements both 
Ruppert's algorithm \cite{Ruppert95} and Chew's algorithm \cite{Chew93},
while the latter implements 
Ruppert's algorithm.
Recently, Chen \textit{et al.} \cite{Chen17} propose a GPU-based algorithm {\tt gQM} that achieves an order of magnitude speed up comparing to {\tt Triangle}.
In the following, we focus our description on our modification to {\tt gQM} to derive our {\tt gDP2d}. Interested reader can refer to  \cite{Chen17} for further technical details. 

\subsection{Overview of {\tt gDP2d}}

Algorithm~\ref{algo_gDP2d} shows the high-level flow of  {\tt gDP2d}.
The main idea is to repeat the iteration ({\tt Line} {\tt 2} to {\tt Line} {\tt 9}) of concurrently subdividing encroached subsegments by their midpoints and breaking up bad triangles by their circumcenters until no further processing needed to improve the quality of the triangulation. Midpoints and circumcenters are termed {\em splitting points} when they are considered for insertion into $\mathcal{T}$ to become its Steiner points. The latter type of Steiner point is also termed  a {\em free point} in $\mathcal{T}$. 
A subsegment $\ell$ is {\em encroached} if there exists a vertex $p$ of $\mathcal{T}$ or a splitting point inside the diametric circle under the Ruppert's algorithm or diametric lenses under the Chew's algorithm.
In this case, we say $p$ {\em encroaches} $\ell$.

\begin{algorithm}[t]
\DontPrintSemicolon
\KwIn{PSLG $\mathcal{G}$; constant $\theta$ and $l$}
\KwOut{Quality Mesh $\mathcal{T}$, which is a constrained or conforming Delaunay triangulation}

Compute the CDT $\mathcal{T}$ of $\mathcal{G}$\;
\Repeat{$L = \emptyset$}
{
    Collect encroached subsegments and bad triangles into $L$ ({\bf Rule~1,~4}) \;
    \If{$L \neq \emptyset$}
    {
        Compute the splitting point set $\mathcal{S}$ of $L$\;
        Prioritize, locate and filter out points in $\mathcal{S}$ ({\bf Rule~2}) \; 
        Filter $\mathcal{S}$ with cavity approximation ({\bf Rule~2,~3})\;
        Insert $\mathcal{S}$ into $\mathcal{T}$ with {\tt Flip-Flop} ({\bf Rule~4}) \; 
    }
}

\caption{\tt gDP2d}
\label{algo_gDP2d}
\end{algorithm}

To ensure termination of the algorithm, each midpoint has priority over circumcenter when considered for insertion, and no two splitting points (midpoint or circumcenters) can be inserted as Steiner points that form an edge in $\mathcal{T}$ in a same iteration. For the former, it means also that a circumcenter inserted into $\mathcal{T}$ as a free point must be undone as it is now {\em redundant} if it encroaches a subsegment where a midpoint of the subsegment is to be inserted instead. For the latter, once the splitting points are known to be connected and they are termed {\em Delaunay dependent}, the one with a lower priority is now redundant and needs to be removed. 
Chen et al. \cite{Chen17} handle such groups of dependent splitting points by processing the splitting points for subsegments before those of bad triangles in turn 
by applying {\tt Flip-Flop} algorithm thrice
\cite{Gao17}.
Their approach resulted in fluctuating workloads.

{\bf Rule 4} is thus one important contributor to performance improvement to our development of {\tt gDP2d} which unifies the processing of encroached subsegments and bad triangles ({\tt Line} {\tt 3} and {\tt Line} {\tt 8}) to use {\tt Flip-Flop} just once. In addition, 
we have incorporated {\bf Rule~1}, {\bf Rule 2} and {\bf Rule~3} too as marked in Algorithm~\ref{algo_gDP2d}. 
Note that the scenario of a serial of low workloads was not noticed significantly in our implementation, thus {\bf Rule 5} is not applicable.

\subsection{One Iteration in {\tt gDP2d}}
Starting with a CDT $\mathcal{T}$ at {\tt Line} {\tt 1} (which can be obtained very fast with, for example, \cite{Shewchuk96} in CPU or \cite{Qi12} in GPU), we process all {\em elements} of encroached subsegments and bad triangles together in one iteration ({\tt Line} {\tt 2} to {\tt Line} {\tt 9}) in accordance to {\bf Rule 4}. They are collected in an array in GPU global memory ({\tt Line} {\tt 3}), and their splitting points are calculated concurrently ({\tt Line} {\tt 5}). When there are few elements $L$ to be processed, the usual compaction for ({\tt Line}~{\tt 3}) can be suppressed in accordance to {\bf Rule~1}. 

For all the splitting points calculated, there are priorities assigned ({\tt Line} {\tt 6}) where midpoint of a longer subsegment has higher priority over that of a shorter one, circumcenter of a larger area triangle has higher priority over that of a smaller one, and all midpoints have priorities over all circumcenters.
A walking process from triangle to triangle in $\mathcal{T}$ is performed concurrently from each element till the triangle of $\mathcal{T}$ containing its splitting point is located ({\tt Line}~{\tt 6}). Note that in the process of locating a splitting point of a bad triangle, we may pass by a triangle incident to some subsegment. In this case, we replace this splitting point by the midpoint of the subsegment as the new splitting point and mark the subsegment as encroached to process it rather than the bad triangle.

Each triangle in $\mathcal{T}$ can be possibly split by at most one splitting point. So, when a triangle is associated with two or more splitting points, these points are Delaunay dependent and only the one with the highest priority will be retained ({\tt Line}~{\tt 6}). This is in accordance to {\bf Rule 2} to reduce the subsequent useless work. 

After {\tt Line}~{\tt 6}, there can still be many pairs of splitting points that are Delaunay dependent but not lying in same triangles of $\mathcal{T}$. Each pair can be identified if the pair was already inserted into $\mathcal{T}$ and was indeed connected for $\mathcal{T}$ to be Delaunay. This is explicit and comprehensive, but messy in that many unnecessary splitting points are simply removed after being added. We design a kernel 
 ({\tt Line} {\tt 7}) to identify for each splitting point $p$, starting from the triangle in $\mathcal{T}$ containing $p$, stepping from triangle to triangle, to include the set ${N}_p$ of some fixed integer $n$ neighboring triangles whose circumcircles enclose $p$. 
Such ${N}_p$ is an approximation of the {\em cavity} of $p$, which is a region formed by the union of those triangles whose circumcircles enclose $p$. Assume a similar ${N}_q$ which is an approximation of the {\em cavity} of another splitting point $q$.  Then, if ${N}_p$ and ${N}_q$ contain a same triangle (that can be found through atomic operations to mark triangles), $p$ and $q$ are not independent and one of them is dropped from subsequent processing in accordance to {\bf Rule 2}. Note that an integer $n$ is used so that this filtering ({\tt Line} {\tt 7}) is of balanced workload for each thread to handle each splitting point to filter out good number, but not all, splitting points in accordance to {\bf Rule 3}. 

The next step ({\tt Line} {\tt 8}) is the actual insertion of splitting points.
It is adapted from {\tt Flip-Flop} in \cite{Chen17} where that insertion is performed in order according to the priorities satisfying the condition that no two splitting points inserted are connected in $\mathcal{T}$ with $\mathcal{T}$ being a CDT.
A {\em flip} is performed to convert two incident triangles $cab$ and $acd$, forming a convex region $abcd$, to the alternate triangles $abd$ and $dbc$. Such a flip reduces the degree of $a$ and $c$, and increases that of $b$ and $d$. A {\em flop} is performed for three triangles mutually incident to each other as $abc$, $acd$, $adb$ to 
remove $a$ to form just triangle $bcd$. So, to remove a (redundant) point $a$, one can perform enough flips to reduce the degree of $a$ till 3 and then do a flop. 

Using flip and flop operations, {\tt Line} {\tt 8} first adds remaining splitting points of $S$ into $\mathcal{T}$ and to repeat the following to reach a CDT of $\mathcal{T}$. 
It identifies and marks redundant points in $\mathcal{T}$ to reduce their degrees with flip till a flop can be performed to remove them from $\mathcal{T}$. Flip is also performed to remove any {\em non-Delaunay edge} where the sum of the two triangle angles opposite the edge is larger than $\pi$. 
After these flips and flops, new redundant points can be discovered for processing, and the process is repeated till $\mathcal{T}$ is a CDT with no redundant points, thus ending the current iteration.

\section{3D Constrained Delaunay}
\label{sec:3D-CDT}

In 3D, a \emph{piecewise linear complex} (PLC) $\mathcal{G}$ consists of a point set $P$, an edge set $E$ (where each edge with endpoints in $P$), and a polygon set $F$ (where each polygon with boundary edges in $E$). A triangulation $\mathcal{T}$ of $\mathcal{G}$ is a decomposition of the convex hull of $P$ into tetrahedra with all points in $P$, edges in $E$ and polygons in $F$ appearing as vertices, union of edges and union of triangles in $\mathcal{T}$. To ease our discussion, we call an edge in $E$ a \emph{segment}, an edge in $\mathcal{T}$ which is also a part or whole of some segment a \emph{subsegment}, and 
a triangle in $\mathcal{T}$ which is also a part or whole of some polygon in $F$ a \emph{subface}. Then, a constrained Delaunay triangulation (CDT) for $\mathcal{G}$ is a triangulation $\mathcal{T}$ such that the circumsphere of each of its tetrahedron $t$ does not contain any vertices except for possibly those not visible (as blocked by polygons in $F$) to all four vertices of $t$.

In a situation where the given PLC includes the watertight boundary of an object, we are interested mainly in the part of the triangulation enclosed within the boundary instead of the convex hull of the vertices of the PLC. 

\subsection{3D CDT Problem}

Given an input PLC $\mathcal{G}$ and a constant $B$, the 3D CDT problem is to compute a CDT $\mathcal{T}$ of $\mathcal{G}$ with few, if not zero, \emph{bad} tetrahedra. A tetrahedron $t$ is {\em bad} if the ratio of the radius of its circumsphere to its shortest edge, termed as \emph{radius-edge ratio}, is larger than $B$. When boundary polygons in $\mathcal{G}$ meet at small angles, any algorithm will end up with some 
bad tetrahedra in the vicinity of the small input angles \cite{Shewchuk14}. 

The 
sequential program {\tt TetGen} \cite{Si15} is the most famous software for this problem. It first computes the CDT $\mathcal{T}$ of the input PLC $\mathcal{G}$, and then iteratively split encroached subsegments, subfaces and bad tetrahedra by inserting Steiner points into $\mathcal{T}$.
A subsegment or subface $e$ is \emph{encroached} when there exists a vertex $p$ of $\mathcal{T}$ or a splitting point inside the diametric sphere of $e$. In this case, we say $p$ is an encroaching point that encroaches $e$.
As in 2D case, {\em splitting points} are points considered for insertion into $\mathcal{T}$ as Steiner points. For an encroached subsegment, we use its midpoint as the splitting point; for a subface or a bad tetrahedron, we use the center of its circumsphere as the splitting point.
To guarantee the termination, the splitting points for mesh elements of lower dimensions have higher priorities than those for mesh elements of higher dimensions.

\begin{algorithm}[t]
\DontPrintSemicolon
\KwIn{PLC $\mathcal{G}$; constant $B$}
\KwOut{Quality mesh $\mathcal{T}$, which is a CDT}
\vspace{0.05in}
Compute the CDT $\mathcal{T}$ of $\mathcal{G}$\;
Split encroached subsegments and subfaces in $\mathcal{T}$ ({\bf Rule~5})\;

\Repeat{$L = \emptyset$}
{
    Collect encroached subsegments, subfaces and bad tetrahedra into $L$ ({\bf Rule~1,~2,~4})\;
    \If{$L \neq \emptyset$}
    {
        Compute the splitting point set $\mathcal{S}$ of $L$\;
        Prioritize and locate the points in $\mathcal{S}$ ({\bf Rule~2})\;
        Grow and shrink cavities of points in $\mathcal{S}$ ({\bf Rule~2,~3})\;
        Insert $\mathcal{S}$ into $\mathcal{T}$\; 
    }
}

\caption{\tt gQM3d}
\label{algo_gQM3d}
\end{algorithm}

\subsection{Overview of {\tt gQM3d}}

Our algorithm {\tt gQM3d}, as shown in Algorithm~\ref{algo_gQM3d},
inserts multiple Steiner points in each iteration ({\tt Line}~{\tt 3} to {\tt Line}~{\tt 10}) until no further processing needed to improve the quality of the triangulation. 
First, it computes the CDT $\mathcal{T}$ of the input PLC $\mathcal{G}$ ({\tt Line} {\tt 1}). For $\mathcal{G}$, we can assume its CDT exists or we use the approach in \cite{Shewchuk98a} to split its edges to guarantee so. This can be done directly in CPU because this is light. Next, {\tt gQM3d} splits encroached subsegments and subfaces on the CPU ({\tt Line}~{\tt 2}). We can choose to do this on the GPU, but our experiments show that this does not have any particular advantage over the CPU approach because such splittings take many rounds of point insertions but each does not have heavy workload to fill GPU capacity (\textbf{Rule~5}).

At {\tt Line}~{\tt 4}, we first collect all encroached subsegments, subfaces and bad tetrahedra into one tuple list $L$ to allow processing of all of them in one go in a kernel ({\bf Rule 4}), and then at {\tt Line} {\tt 6} concurrently compute their splitting points to store into $\mathcal{S}$.  
As mentioned, splitting points of lower dimension elements have higher priorities than higher dimension elements. Between two elements of the same dimension, it is found through our experiments that the element with a larger measure (length for subsegment, area for subface, and volume for bad tetrahedra) should be given a higher priority to achieve a better speedup ({\tt Line}~{\tt 7}). This can be understood as breaking up larger elements early can create more work to better utilize GPU capacity.
Next, the points are located concurrently, where each of these points $s$ is located by starting from its corresponding subsegment, subface or tetrahedron and walking from tetrahedron to tetrahedron, until the tetrahedron $t$ containing $s$ is reached.

For $s$ to be inserted into $\mathcal{T}$ ({\tt Line}~{\tt 9}), we need to compute the set of tetrahedra with their circumspheres enclosing $s$ and all vertices visible to $s$, starting with the tetrahedron $t$ found to contain $s$ in the previous step. This set of tetrahedra is called the {\em cavity of} $s$. When $s$ is to be inserted to be a vertex of $\mathcal{T}$, the cavity of $s$ is to be removed and replaced with tetrahedra with a vertex being $s$ to maintain $\mathcal{T}$ as a CDT. This means that no two splitting points with overlapping cavities can be inserted into $\mathcal{T}$ at the same time. Thus, during the computation of cavities at {\tt Line}~{\tt 8}, some points in $\mathcal{S}$ are filtered out when their cavities are found to be overlapping with existing cavities from splitting points of higher priorities. 
This is implemented by an atomic operation to mark tetrahedra in a cavity.  
We filter out each splitting point that cannot successfully mark a tetrahedron in its cavity or lose any of its marking in the process. Further details on {\tt Line}~{\tt 8} are given in the following subsections. 

At {\tt Line}~{\tt 9}, all the splitting points that remain at this stage can be inserted into $\mathcal T$ with the exception that two nearby splitting points $s_i$ and $s_j$ with cavities sharing a triangle $f$ that is not a subface may be Delaunay dependent, and require a new edge $s_i s_j$ to maintain $\mathcal{T}$ as a CDT. If this edge is too short, the algorithm may not terminate. This is prevented by adding a simple check before any insertions, where if the circumsphere of the tetrahedron formed by $s_i$ and $f$ encloses $s_j$, remove either $s_i$ or $s_j$ (based on their priorities).
After that, we do some housekeeping to maintain correct neighborhood information among new tetrahedra of a cavity and between new tetrahedra with existing tetrahedra incident to the cavity. 

We note that Algorithm~\ref{algo_gQM3d} resembles Algorithm~\ref{algo_gDP2d} except for the flipping (specifically, {\tt Flip-Flop}) to insert splitting points to refine a triangulation, as this method does not work for 3D. Instead, Algorithm~\ref{algo_gQM3d} uses the cavity approach method to insert Steiner points, which employs the tuple-based routine {\tt ExpandList} as discussed next. 

\begin{figure*}[th]
	\centering 
	\includegraphics[scale = 0.35]{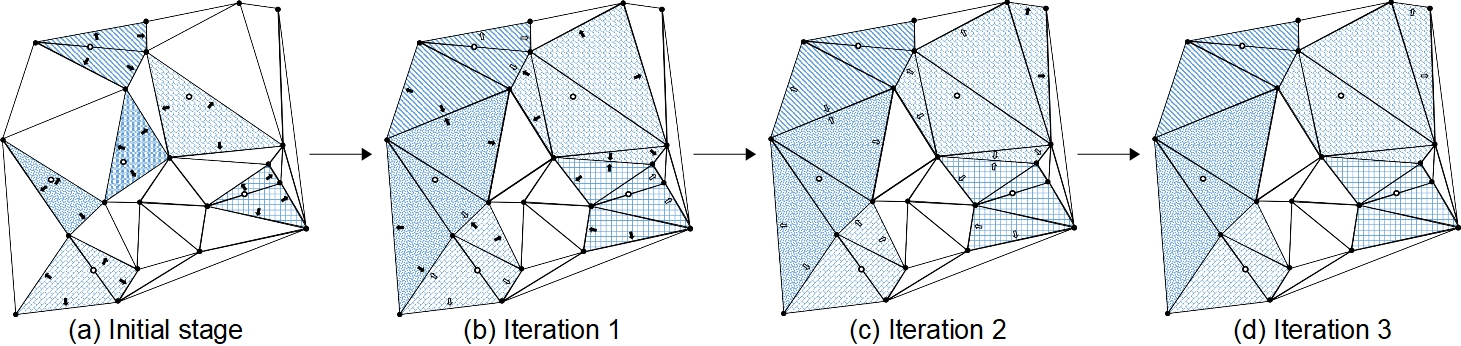}
	\caption{An illustration of the growing of cavities in 2D. 
	(a) Initially, there are 6 splitting points drawn as hollow circles with their initial cavities shown in different grid patterns. A black arrow pointing to an edge $h$ of a triangle $f$ in the cavity of a splitting point $p$ of element $e$ is a tuple $\qdist{e, \qdist{f, h}}$ in $H$. For each tuple, a GPU thread will perform the {\tt incircle} test of $p$ inside the circumcircle of $f'$ where $f'$ shares $h$ with $f$, and attempt to grow the cavity of $p$ by including the tuple $\qdist{e, \qdist{f', h'}}$ and $\qdist{e, \qdist{f', h''}}$ to $H$ where $h', h''$ are the other two edges of $f'$. 
	(b) In iteration 1, the growing of the cavity for one splitting point near the center was unsuccessful due to its presumably lower priority than those of other competing cavities, and this splitting point is removed from the picture. The growing of the others were successful as shown with black arrows as new tuples added to $H$.
	Note that each white arrow indicates the  failure of an {\tt incircle} test for a tuple, and this tuple is added into $\overline{H}$.
	(c) All cavities of splitting points, except for the top-right one, stopped growing.
	(d) As there are no more successful {\tt incircle} test to add new tuples into $H$, the cavity growing is complete.
	}
	\label{fig:growing}
\end{figure*}

\subsection{A Tuple-based {\tt ExpandList} Algorithm}
\label{sec:expandList}

To support the tuple operations required during the computation of cavities at {\tt Line}~{\tt 8}, Algorithm~\ref{algo_gQM3d}, we design the generic Algorithm~\ref{algo_expandlist} that operates on a tuple list $H$. In each iteration, the algorithm performs concurrent operations on a range of tuples (the \textit{operation window}) in the list indicated by {\it left} and {\it right}. Each tuple is first tested with a {\tt Predicate} function. If the test result is positive, an associated {\tt Op\_True} function is performed; otherwise, an {\tt Op\_False} function is performed. Either function can modify existing tuples in $H$ or add a fixed number of new tuples to another tuple list or $H$ itself (reserving positions using prefix sum), thus allowing $H$ and its operation window to expand during processing.

At {\tt Line} {\tt 8}, Algorithm~\ref{algo_expandlist}, the filtering out of tuples in $H$ that need no further work (via standard compaction) can be suppressed if the operation window does not have many tuples. That is, when the system is at low workload we do not need to spend extra effort here just to cause unnecessary lengthening of the latency of {\tt ExpandList} in accordance to {\bf Rule~1}.

\subsection{Growing Cavity}

The cavities for splitting points can have a large variance in size (number of tetrahedra).  Finding a complete cavity in a single GPU thread will result in severely unbalanced work, where there will be many threads which complete their work early forced to wait for the slowest thread.
To workaround this, we partition the work of computing a single cavity further, so that they can be processed by multiple threads, thus reducing the expected variance in latencies of the threads.
This approach works fully with tuples compatible with  {\tt ExpandList} (Algorithm~\ref{algo_expandlist}).
This is in accordance to {\bf Rule~3}. See Figure~\ref{fig:growing} for an illustration of the growing of cavities in 2D.

Initially, for a mesh element $e$ in $L$ (Algorithm~\ref{algo_gQM3d}) with splitting point $p \in \mathcal{S}$ and $\cap(p)$ denoting all tetrahedra in $\mathcal{T}$ intersecting $p$, we add all tuples in $\{\qdist{e, \qdist{t',f}} | f \mbox{ is a facet of } t \in \cap(p)  \mbox{ and } f \cap p = \emptyset  \mbox{ and } t \cap t' = f \}$ into $H$. That is, the cavity of $p$ known initially consists of those tetrahedra $t$ in $\cap(p)$, and we now put all neighbors $t'$ of $t$ into $H$ for checking. 
Here, the {\tt Predicate} is the {\tt insphere} that, working on $h=\qdist{e, \qdist{t',f}}$, returns {\tt true} when the circumsphere of $t'$ encloses $p$ (which is known through $e$); otherwise it returns {\tt false}. 
{\tt Op\_True} is to then attempt to mark $t$ with the priority of $p$ by the atomic operation to include $t$ into its cavity. If succeeded, it continues to add the at most three other neighbors (excluding $t$) of $t'$ into $H$ for the next round of processing; otherwise, $p$ is to filtered out from $\mathcal{S}$.
As for {\tt Op\_False}, it now has found a tetrahedron $t$ immediately outside the cavity of $p$, and it can add $\qdist{e, \qdist{t, f}}$ into $\overline{H}$, which is a list to keep track of all tetrahedra that failed the {\tt insphere} test and will be used subsequently in the shrinking of cavities.

\begin{algorithm}[t]
\DontPrintSemicolon
\KwIn{Tuple list $H$; {\tt Predicate}; {\tt Op\_True}; {\tt Op\_False}}
\KwOut{Augmented tuple list $H$}

Set $\mbox{\it left} = 0$, and $\mbox{\it right} = $ {\tt length}($H$) $- 1$\;
\Repeat{$\mbox{left} > \mbox{right}$}
{
    \For{each tuple $h \in H[\mbox{left} \ldots \mbox{right}]$ }
    {
        \If{{\tt Predicate(}$h${\tt )}}
            {\tt Op\_True($h$)   }
        \Else{\tt Op\_False($h$)  }
    }
    Filter out invalid tuples in $H$ ({\bf Rule 1}) \;
    Set $\mbox{\it left} = \mbox{\it right} + 1$, and $\mbox{\it right} = $ {\tt length}($H$) $- 1$\;
}

\caption{\tt ExpandList}
\label{algo_expandlist}
\end{algorithm}

The above growing approach does not respect visibility so as to detect more encroaching points that are inappropriate for insertion. Thus, we need to shrink the cavities to respect visibility again. To this end, we use {\tt ExpandList} to augment $\overline{H}$ progressively by all tetrahedra belonging to no cavities. Then, the final cavities can be obtained from $H$ and $\overline{H}$ and used for the remeshing purpose. In addition, as subfaces are constraints from the input that we need to split into smaller ones in the output, we also need to identify the \emph{subface cavities}, which can be grown by adding triangles likewise. 
Note that we have simplified the discussion on the growing and shrinking processes, and omitted on some technical details to reach the final Delaunay independent point set. Details can be found in \cite{Chen19}. 

The more bad elements (subsegments, subfaces or tetrahedra) are collected, the longer the list $L$ at {\tt Line}~{\tt 4} (Algorithm~\ref{algo_gQM3d}). This makes high workload a common scenario when growing cavity at {\tt Line}~{\tt 8} (Algorithm~\ref{algo_gQM3d}).
By {\bf Rule~2}, we should attempt to identify dependent work so as to filter them out. Towards this, we have explored using the approximation of the cavity of a splitting point by a 3D grid as a fast test to filter out splitting points that have conflict, similar to the case in 2D. In practice, we also observed that $L$ can become very large without any filtering, which can result in insufficient memory in the GPU. To workaround this issue, we keep $L$ sorted and maintain only the portion with the highest priorities on the GPU. 

\section{3D Restricted Delaunay}
\label{sec:3D-RDT}

In 3D, we call a Delaunay triangulation $\mathcal{T}$ the  {\em restricted Delaunay triangulation} (RDT) with reference to a PLC $\mathcal{G}$, consisting of a point set $P$, an edge set $E$ and a polygon set $F$, when we have the facets of $\mathcal{T}$ {\em identify} with polygons in $F$ of $\mathcal{G}$ as follows. 
The {\em Voronoi dual} of a facet $f$ of some tetrahedron of $\mathcal{T}$ is either (1) a line segment joining the centers of circumspheres of the two tetrahedra (if exist) incident to $f$, or (2) a ray that is perpendicular to $f$ and starts at the center of the circumsphere of the only tetrahedron incident to $f$. 
A facet $f$ {\em identifies} with a polygon $g$ in $F$ if the Voronoi dual of $f$ intersects $g$ (and $g$ is the closest to $f$ among all such polygons in $F$). Such a facet $f$ is also termed a {\em subface}. 
That is, a facet $f$ of a RDT $\mathcal{T}$ is a subface representing the boundary of $\mathcal{G}$ when it can identify with some polygon in $F$ of $\mathcal{G}$; otherwise, it is just one in the interior of $\mathcal{T}$. A quality RDT with reference to $\mathcal{G}$ is to represent $\mathcal{G}$ for use in applications.    

To simplify our discussion, we assume polygons in $F$ of $\mathcal{G}$ form a water-tight 3D volume. In this case, a RDT $\mathcal{T}$ with reference to $\mathcal{G}$ is a 3D simplex formed by a collection of tetrahedra that defines the interior or volume of $\mathcal{T}$.
Our discussion here can be applied to general $\mathcal{G}$ that is not water-tight. Examples of such problem sets are included in our  experiments (Section~\ref{sec:exp3D-RDT}).

\subsection{3D RDT Problem}
Given an input PLC $\mathcal{G}$, and constants $B$, $R$, $\theta$, $r$, the 3D RDT problem is to compute a RDT $\mathcal{T}$ that contains no bad subfaces and no bad tetrahedra with reference to $\mathcal{G}$. 
A subface $f$ is {\em bad} if its minimum angle exceeds $\theta$ or the radius of its {Delaunay sphere} is larger than $r$. 
The {\em Delaunay sphere} of $f$ is the sphere passing through vertices of $f$ and having center at
the intersection point of the Voronoi dual of $f$ with the polygon in $F$ of $\mathcal{G}$ identified with $f$.
A tetrahedron $t$ is {\em bad} if its radius-edge ratio is larger than $B$, or the radius of its circumsphere is larger than $R$. 

An algorithm for the problem, such as \cite{Boissonnat05, Cheng08}, is to refine a bad subface or a bad tetrahedron in each iteration by inserting (or re-sampling) Steiner points into $\mathcal{T}$, and to terminate when there is no more bad subface nor bad tetrahedron.
Once again, {\em splitting points} are points considered for insertion into $\mathcal{T}$ as Steiner points. For a bad subface $f$, we use the center of its Delaunay sphere as a splitting point; for a bad tetrahedron, we use the circumcenter of its circumsphere.    
The splitting points for subfaces have higher priorities for insertion than those for bad tetrahedra.
In the process of insertion, a subface may subsequently lie within $\mathcal{T}$ as an interior facets and no longer a subface on the boundary, and other new facets (of new tetrahedra added) may become subfaces of $\mathcal{T}$.
The 
{\tt CGAL} 3D mesh generation package \cite{cgal3d, Jamin15} 
is also based on this iterative approach.

\begin{algorithm}[t]
\DontPrintSemicolon
\KwIn{PLC $\mathcal{G}$; constants $B$, $R$, $\theta$ and $r$}
\KwOut{Restricted Delaunay triangulation $\mathcal{T}$, with no bad subface nor bad tetrahedra, with reference to $\mathcal{G}$}

Compute a $\mathcal{T}$ with a small number of vertices in $P$ of $\mathcal{G}$\;
\Repeat{$L = \emptyset$}
{
    Collect bad subfaces, bad tetrahedra into $L$ ({\bf Rule~1,~2,~4})\;
    \If{$L \neq \emptyset$}
    {
        Compute the splitting point set $\mathcal{S}$ of $L$\;
        Prioritize and locate the points in $\mathcal{S}$ ({\bf Rule~2})\;
        Compute cavities of points in $\mathcal{S}$ ({\bf Rule~2,~3,~5})\;
        Insert $\mathcal{S}$ into $\mathcal{T}$\; 
        Update status of facets and tetrahedra ({\bf Rule~2,~3,~4})\;
    }
}

\caption{\tt gDP3d}
\label{algo_gDP3d}
\end{algorithm}

\subsection{Overview of {\tt gDP3d}}

Our algorithm {\tt gDP3d}, as shown in Algorithm~\ref{algo_gDP3d}, inserts multiple Steiner points in each iteration ({\tt Line}~{\tt 2} to {\tt Line}~{\tt 10}) until no bad subface or bad tetrahedron remains. 
First, it computes a small Delaunay triangulation $\mathcal{T}$ ({\tt Line} {\tt 1}) from a random sampling of some vertices in $P$ of $\mathcal{G}$. This can be done directly in GPU or in CPU because this is light. 
To assist our computation, $\mathcal{T}$ is placed in a big bounding box containing all elements of $\mathcal{G}$, and the bounding box is triangulated with tetrahedra with those forming a part of $\mathcal{T}$ as representing the interior of $\mathcal{T}$ while the other the exterior to $\mathcal{T}$.

We note that Algorithm~\ref{algo_gDP3d} resembles Algorithm~\ref{algo_gQM3d}, but uses the {\tt ExpandList} for not only growing cavities but also testing intersection, both of which are described as follows.

To compute the cavities, unlike Algorithm~\ref{algo_gQM3d} who uses grow-and-shrink scheme, {\tt gDP3d} only needs to grow the cavities because the RDT $\mathcal{T}$ is always a Delaunay triangulation ({\tt Line}~{\tt 7}). However, this means some splitting points can have huge cavities. 
Although {\tt ExpandList} is designed to promote work balancing among GPU threads when growing the cavity, it is still possible to reach a case when most, but not all, growing are done and the remaining threads continue to working on just a few tuples in $H$ over many more rounds ({\tt Line}~{\tt 2} to {\tt Line}~{\tt 10}, Algorithm~\ref{algo_expandlist}).
Thus, we have 
low workload of {\tt Op\_True} and {\tt Op\_False} over many iterations. To address this waste, 
in accordance to {\bf Rule~5}, we choose to record those partial cavities appearing in $H$ when the length of $H$ becomes short and pass these information to be a part of the list $L$ in Algorithm~\ref{algo_gDP3d} ({\tt Line}~{\tt 3}). Such partial cavities are then to be accumulated and completed with others in subsequent iterations.     

With more tetrahedra added to $\mathcal{T}$, it gradually approaches $\mathcal{G}$. At {\tt Line} {\tt 9}, we remove all existing subfaces and mark new subfaces among facets of new tetrahedra when their Voronoi duals intersect some polygons in $F$ of $\mathcal{G}$.
In addition, we need to mark new tetrahedra created at cavities as being interior or exterior to $\mathcal{T}$. Interior tetrahedra are part of the desired output and they should not be bad tetrahedra. Exterior tetrahedra have no such restrictions and are not considered as candidates for subsequent refinement, but may still be replaced during the process of further refinement. Further details on {\tt Line}~{\tt 9} are in the following subsection.

\subsection{Testing Intersection}

With new points in $\mathcal{T}$ that form new facets, we need to process all newly created facets and existing subfaces to identify which ones are (still) subfaces. For each one, it is a subface when the line segment representing its Voronoi dual intersects some polygon in $F$ of $\mathcal{G}$.  
As for newly created tetrahedra, we need to differentiate interior from exterior ones. For a tetrahedron $t$, we take a random ray starting from its interior and shooting out of $t$ till some point on the surface of the bounding box of $\mathcal{G}$ to count the number of intersections of this ray with polygons in $F$. If the count is odd, then $t$ is an interior tetrahedron; otherwise, $t$ is an exterior one. 

Both the above-mentioned operations on processing facets and tetrahedra involve intersection tests of line segments with polygons in $F$ in GPU. The former is usually with short line segments, while the latter on long ones (rays). To support this, we build an AABB (Axis-Aligned Bounding Box) tree for the polygons in $F$, and then fatten this balanced binary tree in an array to be stored in the global memory in GPU. The root node is the bounding box of all polygons of $F$. At each node, we have the bounding box of those polygons represented by the node, and then polygons are sorted along the longest axis of the bounding box and divided in the middle to be stored on the left and on the right child. Note that we take the simple approach of storing a polygon in just one child if it is cut by the dividing plane. Each leaf node contains one polygon of $F$. Note that the AABB tree can be computed very efficiently on the CPU, then subsequently moved to the GPU to perform the intersection tests. To find an intersection of a line segment or a ray $r$ with a node representing some polygons, we check whether $r$ intersects the bounding box of the node. If yes, recursively check with the left and the right child till reaching the leaf nodes where $r$ is tested for any intersection with a polygon stored with each leaf.
In practice, the height of the tree is bounded by a small constant ($< 20$ for our case) and all intersection tests needed can be done very quickly with AABB tree as follows.

{\tt Line} {\tt 8} of Algorithm~\ref{algo_gDP3d} creates new tetrahedra in cavities of splitting points inserted into $\mathcal{T}$. Each new tetrahedron contributes four segments of Voronoi duals of facets for intersection tests, and a ray from itself towards the bounding box of $\mathcal{G}$ for counting number of intersections with polygons in $F$. 
We can re-fit Algorithm~\ref{algo_expandlist} to do intersection tests concurrently for all these segments (or rays).
Initially, we have $H$ to store tuples where each $\qdist{\ell, \mbox{\it node}}$ represents a segment $\ell$ (or ray) and $\mbox{\it node}$ is the root node of the AABB tree to be tested for intersection. The {\tt Predicate} is to test whether the tuple $\qdist{\ell, \mbox{\it node}}$ is such that $\ell$ intersects $\mbox{\it node}$. Then, {\tt Op\_True} for $h=\qdist{\ell, \mbox{\it node}}$ is to add the two children of $\mbox{\it node}$ in the AABB tree to $H$ while {\tt Op\_False} marks the tuple to invalid for no subsequent processing needed.    
When {\tt ExpandList} completes, each tuple in $H$ represents a pair of intersection between $\ell$ and one polygon in $F$. For $\ell$ which is a Voronoi dual of a facet $f$, this facet is then marked as a subface. For $\ell$ which is a ray out of a tetrahedron $t$, we count the total (valid) occurrences of $\ell$ appearing in $H$ to determine that $t$ is interior to $\mathcal{T}$ if the count is odd, and exterior if even. And this step is complete.

We now review the above approach with reference to our design rules. With the AABB tree, we could actually use one thread for each element to traverse the tree to find or count intersections. This method suffers from unbalanced work because the number of intersection is unknown. Our approach with {\tt ExpandList} to vectorize work into a tuple list $H$ is in accordance to {\bf Rule 3}. As such, the computation is often at the good case of high workload scenarios.

To further improve the performance, we want to control the high workload in accordance to {\bf Rule 2} still. So, for deciding whether a facet $f$ is a subface, we use a shortcut to check possible intersection of the Voronoi dual of $f$ with those polygons in $F$ containing the vertices of $f$. If found, $f$ is then a subface and its splitting point is (any one of) the intersection point. 
If not found, then proceed as mentioned before to find intersection through the AABB tree. 

Next, {\bf Rule 4} appears in the consideration of {\tt Line} {\tt 3} and {\tt Line}~{\tt 9} of Algorithm~\ref{algo_gDP3d}. In particular, there are two groups of elements, bad subfaces and bad tetrahedra, to be refined with the former of higher priority to the latter. The above discussion assumes that both are processed concurrently respecting their priorities. That is, we have unified the processing of two groups into one. But, the  intersection test for the former needs only a cheaper {\tt true} or {\tt false} answer rather than, for the latter, a more expensive odd-even count to mean an interior or exterior tetrahedron. Putting both together as a same routine of just an odd-even count does result in a higher cost for the former but may result in a better overall throughput ({\bf Rule~4}).

From our experiments with the implementation of {\tt gDP3d}, we note there are already enough facets to work with when the subfaces of $\mathcal{T}$ do not yet form a good approximation of the boundary of $\mathcal{G}$. In this case, the unifying is actually a penalty to the performance because the throughput becomes low with the increase in the latency with odd-even intersections count for many facets and a relatively small number of tetrahedra. It is when $\mathcal{T}$ has subfaces closely approximating the boundary of $\mathcal{G}$ that the refinement shifts to refining many interior tetrahedra of $\mathcal{T}$ and a relatively small number of subfaces. For this, the unifying has better throughput because the higher cost for intersection test is anyway needed for new tetrahedra (excluding those split from interior tetrahedra that are known to be interior without any intersection test), whereas the relatively small number of facets are just piggybacking their intersection tests. This avoids the need to do intersection tests for facets separately as a kernel at low workload with waste to GPU capability.

\section{Experimental Results}
\label{sec:experiment}

All experiments were conducted on a PC with an Intel i7-7700k, 4.2GHz, 4 Cores, 8 Threads CPU, 32GB of DDR4 RAM and a GTX1080 Ti graphics card with 11GB of video memory. All software was complied with all optimization flags turned on. We implemented {\tt gDP2d}, {\tt gQM3d} and {\tt gDP3d} using the CUDA programming model of NVIDIA \cite{nvidia:cuda}. We use the exact arithmetic and robust geometric predicate of Shewchuk \cite{Shewchuk97b} to handle robustness and the simulation of simplicity \cite{Edelsbrunner90} to deal with degenerate cases. 
For each input, we report the average timing and triangulation quality over multiple runs.


\subsection{2D Constrained Delaunay Refinement}

{\tt gDP2d}, as implemented in accordance to the design rules, shows in general 50\% improvement in speed up compared to our prior work of {\tt gQM} \cite{Chen17}. 
We further compare {\tt gDP2d} with {\tt CGAL} 2D mesh generator \cite{cgal2d} and {\tt Triangle} software. We use running time and output triangulation quality of {\tt Triangle} as the base case. In measuring computing time, we exclude the time of computing the CDT of the input PSLG ({\tt Line} {\tt 1} of Algorithm~\ref{algo_gDP2d}).
Extensive experimental results on synthetic datasets for the various software are available in \cite{Chen:Thesis}. In sum, {\tt gDP2d} is an order or two magnitude faster than {\tt CGAL} and {\tt Triangle} while inserting comparable amounts of Steiner points.

As for real world datasets, Table~\ref{table:2d_realworld} shows our findings for the software on contour maps freely available at http://www.ga.gov.au/  under Ruppert's mode (and similar findings under Chew's mode and thus omitted here). We set $\theta = 20^\circ$ to guarantee that all software can terminate \cite{Ruppert95}. To further stress test them, we extend {\tt Triangle} to support the edge length criterion $l$, and set $l$ to small values to exhaust GPU memory. Note that real world datasets usually have some small input angles and thus it is unavoidable to have some bad triangles in the output triangulations. So, 
Table~\ref{table:2d_realworld} records, besides the number of output points (third column) and time to complete the computation (fifth column), the sum of bad areas that are occupied by resulting bad triangles (fourth column). In sum, although {\tt CGAL} inserts fewer points, it over-protects the vicinity of small input angles (to have no Steiner points in those places which leads to very large bad area) and its outputs may not be good for applications. 
The speed up of {\tt gDP2d} over {\tt Triangle} can reach up to 16 times on the 8.4M sample. On the other hand, {\tt gDP2d} inserts slightly more Steiner points than {\tt Triangle} when the input sizes are large. 

\begin{table}[!t]
    \centering
    \scalebox{1.0}
    {
    \begin{tabular}{c|c|c c c c}
         \hline
         \multicolumn{1}{r|}{} & & & {\tt Bad} & & \\
         \multicolumn{1}{r|}{\tt Name} & {\tt Software} & {\tt Points} & {\tt Area} & {\tt Time} & {\tt Speed}  \\
         \multicolumn{1}{r|}{$l$} & & {\tt (M)} & {\tt (\%)} & {\tt (s)} & {\tt Up}\\
         \hline
         \multicolumn{1}{r|}{} & {\tt Triangle} & 35.4 & 0 & 67 & 1 \\
         \multicolumn{1}{r|}{1.2M} & {\tt CGAL} & 4.1 & 91.1 & 46 & - \\
         \multicolumn{1}{r|}{0.12} & {\tt gDP2d} & 34.1 & 0 & 14 & 5 \\
         \hline
         \multicolumn{1}{r|}{} & {\tt Triangle} & 34.1 & 0 & 114 & 1 \\
         \multicolumn{1}{r|}{3.2M} & {\tt CGAL} & 5.8 & 92.1 & 46 & - \\
         \multicolumn{1}{r|}{0.12} & {\tt gDP2d} & 33.5 & 0 & 15 & 8 \\
         \hline
         \multicolumn{1}{r|}{} & {\tt Triangle} & 30.3 & 0 & 116 & 1 \\
         \multicolumn{1}{r|}{4.3M} & {\tt CGAL} & 9.0 & 88.7 & 65 & - \\
         \multicolumn{1}{r|}{0.11} & {\tt gDP2d} & 30.6 & 0 & 13 & 9 \\
         \hline
          \multicolumn{1}{r|}{} & {\tt Triangle} & 30.4 & 0 & 133 & 1 \\
         \multicolumn{1}{r|}{5.6M} & {\tt CGAL} & 8.7 & 91.6 & 59 & - \\
         \multicolumn{1}{r|}{0.1} & {\tt gDP2d} & 31.5 & 0 & 16 & 8 \\
         \hline
         \multicolumn{1}{r|}{} & {\tt Triangle} & 26.6 & 0 & 251 & 1 \\
         \multicolumn{1}{r|}{8.4M} & {\tt CGAL} & 12.5 & 87.8 & 82 & - \\
         \multicolumn{1}{r|}{0.12} & {\tt gDP2d} & 29.2 & 0 & 16 & 16 \\
         \hline
         \multicolumn{1}{r|}{} & {\tt Triangle} & 29.3 & 0 & 345 & 1 \\
         \multicolumn{1}{r|}{9.4M} & {\tt CGAL} & 13.9 & 88.5 & 93 & - \\
         \multicolumn{1}{r|}{0.16} & {\tt gDP2d} & 32.0 & 0 & 25 & 14 \\
         \hline
    \end{tabular}
    }
   \vspace{5pt}
    \caption{2D CDT experimental results on some contour maps.}
    \label{table:2d_realworld}
\end{table}


\begin{table}[!t]
    \centering
    \scalebox{1.0}
    {
    \begin{tabular}{c|c|c c c c c}
         \hline
         \multicolumn{1}{c|}{\tt } & \multirow{3}{*}{\tt Software} & 
         & {\tt Bad} & &\\
         \multicolumn{1}{c|}{\tt Model ID} & & {\tt Points} & {\tt Tets} & {\tt Time} & {\tt Speed}\\
         \multicolumn{1}{c|}{\tt Name} & & {\tt (M)} & {\tt (M)} & {\tt (min)} & {\tt Up}\\
         \hline
         \multicolumn{1}{c|}{63788} & {\tt TetGen} & 4.27 & 1.63 & 66.5 & 1 \\
         \multicolumn{1}{c|}{Skull} & {\tt gQM3d} & 4.23 & 1.67 & 0.7 & 102\\
         \hline
         \multicolumn{1}{c|}{65942} & {\tt TetGen} & 4.54 & 1.87 & 71.4 & 1 \\
         \multicolumn{1}{c|}{Sculpture} & {\tt gQM3d} & 4.50 & 1.87 & 1.0 & 69\\
         \hline
         \multicolumn{1}{c|}{63785} & {\tt TetGen} & 3.29 & 1.23 & 39.3 & 1 \\
         \multicolumn{1}{c|}{Half-skull} & {\tt gQM3d} & 3.37 & 1.23 & 0.8 & 52\\
         \hline
         \multicolumn{1}{c|}{94059} & {\tt TetGen} & 2.54 & 1.35 & 24.8 & 1 \\
         \multicolumn{1}{c|}{Mask} & {\tt gQM3d} & 2.54 & 1.35 & 0.5 & 50\\
         \hline
         \multicolumn{1}{c|}{1717685} & {\tt TetGen} & 2.98 & 0.90 & 24.6 & 1\\
         \multicolumn{1}{c|}{Brain} & {\tt gQM3d} & 2.99 & 0.90 & 0.9 & 28\\
         \hline
         \multicolumn{1}{c|}{793565} & {\tt TetGen} & 2.13 & 0.93 & 17.2 & 1\\
         \multicolumn{1}{c|}{SkullBox} & {\tt gQM3d} & 2.16 & 0.93 & 0.6 & 28\\
         \hline
         \multicolumn{1}{c|}{461112} & {\tt TetGen} & 3.29 & 1.27 & 28.9 & 1 \\
         \multicolumn{1}{c|}{Mutant} & {\tt gQM3d} & 3.40 & 1.27 & 1.0 & 28\\
         \hline
         \multicolumn{1}{c|}{763718} & {\tt TetGen} & 1.83 & 0.83 & 12.1 & 1 \\
         \multicolumn{1}{c|}{Jacket} & {\tt gQM3d} & 1.79 & 0.83 & 0.5 & 24\\
         \hline
         \multicolumn{1}{c|}{252653} & {\tt TetGen} & 2.93 & 1.04 & 29.5 & 1 \\
         \multicolumn{1}{c|}{Thunder} & {\tt gQM3d} & 2.94 & 1.04 & 1.6 & 19\\
         \hline
         \multicolumn{1}{c|}{87688} & {\tt TetGen} & 1.20 & 0.61 & 5.4 & 1 \\
         \multicolumn{1}{c|}{Shy-light} & {\tt gQM3d} & 1.21 & 0.62 & 0.6 & 9\\
         \hline
    \end{tabular}
    }
    \vspace{5pt}
    \caption{3D CDT Experimental results on some samples in Thingi10k.}
    \label{table:3d-cdt-readworld}
    \vspace{-0.22in}
\end{table}

\subsection{3D Constrained Delaunay Refinement}

{\tt gQM3d}, as implemented in accordance to the design rules, doubles the performance of our prior work in \cite{Chen19}.
On synthetic datasets, {\tt gQM3d} reaches more than 40 times speed up over {\tt TetGen} \cite{Si15}  
while inserting up to 4\% more
Steiner points than {\tt TetGen} (details are available in \cite{Chen:Thesis}).
The running time of {\tt gQM3d} includes the time for computation plus the time needed to transfer data between CPU and GPU.

We also tested {\tt gQM3d} and {\tt TetGen} on some samples from the Thingi10k dataset \cite{Zhou16} with $B = 1.4$ and the results are shown in Table~\ref{table:3d-cdt-readworld}. The speedup of {\tt gQM3d} over {\tt TetGen} reached up to 102 times (sixth column), while producing triangulations with similar sizes (third column) and having similar profile of dihedral angles for their tetrahedra.
Note that both software generated large numbers of bad tetrahedra (fourth column) in the outputs of these real-world samples. This is because the real-world samples have a lot more small input angles which become part of those bad tetrahedra in the outputs. 

Figure \ref{fig:teaser}(a) shows the quality mesh of the Model 763718, Jacket as generated by {\tt gQM3d}, and Figure \ref{fig:teaser}(b) shows the cut-off view of the mesh.


\begin{table}[!t]
    \centering
    \scalebox{1.0}
    {
    \begin{tabular}{c|c|c c c}
         \hline
         \multicolumn{1}{c|}{\tt Model ID} & \multirow{3}{*}{\tt Software} & 
         & &\\
         \multicolumn{1}{c|}{\tt Name} & & {\tt Points}  & {\tt Time} & {\tt Speed} \\
         \multicolumn{1}{c|}{$r$} & & {\tt (M)} & {\tt (min)} & {\tt Up} \\
         \hline
         \multicolumn{1}{c|}{1706475} & {\tt gDP3d} & 6.62 & 0.7 & 1\\
         \multicolumn{1}{c|}{Accessories} & {\tt CGAL-S} & 6.60 & 17.5 & 26\\
         \multicolumn{1}{c|}{0.08} & {\tt CGAL-M} & 6.75 & 45.6 & 67\\
         \hline
         \multicolumn{1}{c|}{45939} & {\tt gDP3d} & 5.86 & 0.7 & 1\\
         \multicolumn{1}{c|}{Gandhi Litho} & {\tt CGAL-S} & 5.84 & 12.0 & 18\\
         \multicolumn{1}{c|}{0.1}  & {\tt CGAL-M} & 5.98 & 40.8 & 60\\
         \hline
         \multicolumn{1}{c|}{236922} & {\tt gDP3d} & 5.91 & 0.6 & 1\\
         \multicolumn{1}{c|}{Aztec} & {\tt CGAL-S} & 5.89 & 15.4 & 27\\
         \multicolumn{1}{c|}{0.05} & {\tt CGAL-M} & 6.03 & 13.6 & 24\\
         \hline
         \multicolumn{1}{c|}{551021} & {\tt gDP3d} & 5.37 & 0.6 & 1\\
         \multicolumn{1}{c|}{Arc Triomphe} & {\tt CGAL-S} & 5.35 & 14.5 & 25\\
         \multicolumn{1}{c|}{0.03} & {\tt CGAL-M} & 5.48 & 10.6 & 18\\
         \hline
         \multicolumn{1}{c|}{65942} & {\tt gDP3d} & 7.37 & 0.8 & 1\\
         \multicolumn{1}{c|}{Sculpture} & {\tt CGAL-S} & 7.35 & 20.5 & 25\\
         \multicolumn{1}{c|}{0.04} & {\tt CGAL-M} & 7.53 & 14.8 & 18\\
         \hline
         \multicolumn{1}{c|}{1088281} & {\tt gDP3d} & 8.34 & 11.0 & 1\\
         \multicolumn{1}{c|}{Letter Z}  & {\tt CGAL-S} & 8.32 & 449.5 & 41\\
         \multicolumn{1}{c|}{0.03} & {\tt CGAL-M} & 8.52 & 179.0 & 16\\
         \hline
         \multicolumn{1}{c|}{1255206} & {\tt gDP3d} & 6.44 & 1.5 & 1\\
         \multicolumn{1}{c|}{Hendecahedron} & {\tt CGAL-S} & 6.42 & 54.4 & 36\\
         \multicolumn{1}{c|}{0.02} & {\tt CGAL-M} & 6.57 & 16.7 & 11\\
         \hline
         \hline
    \end{tabular}
    }
    \vspace{5pt}
    \caption{3D RDT experimental results on some samples in Thingi10k for refining subfaces.}
    \label{table:3d-rdt-sr}
\end{table}

\begin{table}
    \centering
    \scalebox{1.0}
    {
    \begin{tabular}{c|c|c c c}
         \hline
         \multicolumn{1}{c|}{\tt Model ID} & \multirow{3}{*}{\tt Software} & 
         & &\\
         \multicolumn{1}{c|}{\tt Name} & & {\tt Points}  & {\tt Time} & {\tt Speed} \\
         \multicolumn{1}{c|}{$R$} & & {\tt (M)} & {\tt (min)} & {\tt Up} \\
         \hline
         \multicolumn{1}{c|}{1706475} & {\tt gDP3d} & 7.32 & 0.7 & 1\\
         \multicolumn{1}{c|}{Accessories} & {\tt CGAL-S} & 7.23 & 12.0 & 17\\
         \multicolumn{1}{c|}{0.3} & {\tt CGAL-M} & 7.60 & 3.8 & 5\\
         \hline
         \multicolumn{1}{c|}{551021} & {\tt gDP3d} & 6.40 & 0.6 & 1\\
         \multicolumn{1}{c|}{Arc Triomphe} & {\tt CGAL-S} & 6.33 & 10.4 & 19\\
         \multicolumn{1}{c|}{0.13} & {\tt CGAL-M} & 6.65 & 1.2 & 2\\
         \hline
         \multicolumn{1}{c|}{65942} & {\tt gDP3d} & 7.78 & 0.8 & 1\\
         \multicolumn{1}{c|}{Sculpture} & {\tt CGAL-S} & 7.69 & 13.7 & 18\\
         \multicolumn{1}{c|}{0.13} & {\tt CGAL-M} & 8.07 & 1.8 & 2\\
         \hline
         \multicolumn{1}{c|}{1088281} & {\tt gDP3d} & 6.85 & 2.9 & 1\\
         \multicolumn{1}{c|}{Letter Z} & {\tt CGAL-S} & 6.76 & 80.1 & 27\\
         \multicolumn{1}{c|}{0.09} & {\tt CGAL-M} & 7.09 & 40.5 & 14\\
         \hline
         \multicolumn{1}{c|}{1255206} & {\tt gDP3d} & 6.54 & 0.6 & 1\\
         \multicolumn{1}{c|}{Hendecahedron} & {\tt CGAL-S} & 6.47 & 11.5 & 20\\
         \multicolumn{1}{c|}{0.12} & {\tt CGAL-M} & 6.80 & 1.2 & 2\\
         \hline
         \multicolumn{1}{c|}{518031} & {\tt gDP3d} & 7.19 & 0.6 & 1\\
         \multicolumn{1}{c|}{Lampan Hack} & {\tt CGAL-S} & 7.11 & 11.7 & 20\\
         \multicolumn{1}{c|}{0.65} & {\tt CGAL-M} & 7.47 & 1.1 & 2\\
         \hline
         \multicolumn{1}{c|}{940414} & {\tt gDP3d} & 6.50 & 1.9 & 1\\
         \multicolumn{1}{c|}{Voronoi Lamp} & {\tt CGAL-S} & 6.43 & 36.3 & 19\\
         \multicolumn{1}{c|}{0.63} & {\tt CGAL-M} & 6.73 & 7.0 & 4\\
         \hline
    \end{tabular}
    }
    \vspace{5pt}
    \caption{3D RDT experimental results on some samples in Thingi10k for refining interior volume. Gandhi Litho and Aztec in Table~\ref{table:3d-rdt-sr} are not water-tight with interior and thus do not appear in the above.}
    \label{table:3d-rdt-tr}
\end{table}

\subsection{3D Restricted Delaunay Refinement}
\label{sec:exp3D-RDT}

We compare {\tt gDP3d} to {\tt CGAL} 3D mesh generator, including {\tt CGAL-S} (single-threading CPU), {\tt CGAL-M} (multi-threading CPU). Two kinds of tests are done. The first test on refining subfaces sets $\theta = 30^\circ$ to guarantee the termination \cite{Chew93} and $r$ to small values to exhaust GPU memory, while the second test on refining interior volume (on only water-tight model) sets, besides $\theta = 30^\circ$, $B = 2$ to guarantee the termination \cite{Shewchuk98b} and $R$ to small values to exhaust GPU memory. 

We tested all software on some samples from the Thingi10k dataset \cite{Zhou16}. For the first test, the speed up of {\tt gDP3d} over {\tt CGAL-S} and {\tt CGAL-M} reaches up to 41 and 67 times respectively; see Table~\ref{table:3d-rdt-sr}. For the second test, {\tt gDP3d} is also an order of magnitude faster than {\tt CGAL-S} (with 27 times speed up) and {\tt CGAL-M} (with 14 times speed up); see Table~\ref{table:3d-rdt-tr}. 
%
%
Furthermore, {\tt gDP3d} inserts no more than 1.3\% Steiner points compared to {\tt CGAL-S}, and 1.7\% less than {\tt CGAL-M}. 
Besides with good speed up, {\tt gDP3d} produces quality output of a reasonable size. This is also true for other real world samples we have tested on. 
See Figure~\ref{fig:teaser}(c) for the quality mesh of the Model 940414, Voronoi Lamp as generated by {\tt gDP3d}, and Figure~\ref{fig:teaser}(d) shows the cut-off view of the mesh.

\begin{figure}[!t]
    \centering
    \includegraphics[scale = 0.75]{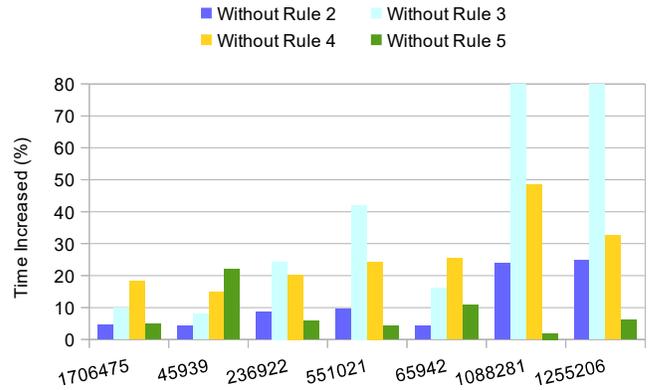}
    \vspace{-0.2in}
    \caption{Increase in running time (in percentage) of {\tt gDP3d} without each of the rules for the real world samples in Table~\ref{table:3d-rdt-sr}. The impact of {\bf Rule~1} is not as significant as the other and thus omitted here.}
    \label{fig:rules_3d}
\end{figure}

We have {\tt gDP3d} achieving its good performance because its implementation is in accordance to the design rules. To show this fact for
{\tt gDP3d} (and similarly can be done for {\tt gDP2d} and {\tt gQM3d} and thus omitted here), we have Figure~\ref{fig:rules_3d} shows the increase in running time (in percentage) of {\tt gDP3d} when we do not incorporate computational components suggested by \textbf{Rule~2} (reduce waste by filtering out some splitting points, and shortcut to test intersection), \textbf{Rule~3} ({\tt ExpandList} instead of one thread to traverse the whole AABB tree), \textbf{Rule~4} (does not unify intersection tests of facets and tetrahedra), and \textbf{Rule~5} (carry forward the computation of large cavities). {\tt gDP3d} is up to 24\% slower without \textbf{Rule~2}, 200\% slower without \textbf{Rule~3}, 48\% slower without \textbf{Rule~4}, and 22\% slower without \textbf{Rule~5}.
\textbf{Rule~3} and \textbf{Rule~4} stand out among all in improving the performance of {\tt gDP3d}. 
The results for the other real world samples and the second test are similar. 

\section{Concluding Remarks}
\label{sec:conclusion}

This paper focuses on the management of waste in the GPU hardware when running computational programs.
It provides a new angle towards analyzing GPU algorithms outside traditional methods such as using total work processed. It presents a set of high-level rules (or strategies) catered to expected scenarios based on workload for a single kernel as well as special cases of workloads of a serial of kernels within a small window to help in the design and development of a performant GPU program.

As shown in the case studies, the rules provided effective guidance in the design of 2D and 3D mesh refinement algorithms {\tt gDP2d}, {\tt gQM3d} and {\tt gDP3d}, respectively, achieving good speed ups. All source codes will be made available online in due course. The rules are formulated over a simple partitioning of cases based on workload or a serial of workloads, and is general enough to be applied to many other applications, though the actual implementation will need to be specialized to the type of problem.

Moving forward, we will explore the applicability of these rules on algorithms for other problems, including, but not limited to, other meshing problems 
\cite{ChengBook2002Meshing}.



\ifCLASSOPTIONcompsoc
  \section*{Acknowledgments}
\else
  \section*{Acknowledgment}
\fi

This research is supported by the National University of Singapore under grant R-252-000-678-133.

\ifCLASSOPTIONcaptionsoff
  \newpage
\fi

\end{document}